\documentclass[aps,pre,showpacs,superscriptaddress,longbibliography,twocolumn,floatfix]{revtex4-1}
\usepackage{graphicx}
\usepackage{amsmath}
\usepackage{amssymb}
\usepackage{mathtools}
\usepackage{makeidx}
\usepackage{amsfonts}
\usepackage{bm}
\usepackage{dcolumn}
\usepackage{color}
\usepackage{xcolor}
\usepackage{xfrac}

\newcommand{\be}{\begin{equation}}
\newcommand{\ee}{\end{equation}}

\begin{document}

\title{Perturbative Steady States of Completely Positive Quantum Master Equations}
\author{Jae Sung Lee}\email{jslee@kias.re.kr}
\affiliation{School of Physics, Korea Institute for Advanced Study, Seoul 02455, Korea}
\author{Joonhyun Yeo} \email{jhyeo@konkuk.ac.kr}
\affiliation{Department of Physics, Konkuk University, Seoul 05029, Korea}

\date{\today}

%%%%%%%%%%%%%%%%%%%%%%%%%%%%%%%%%%%%%%%%%%%%%%%%%%%%%%%%%%%%%%%%%%%%%
\begin{abstract}
The Lindblad form guarantees complete positivity of a Markovian quantum master equation (QME). However, its microscopic derivation for a quantum system weakly interacting with a thermal bath requires several approximations, which may result in inaccuracies in the QME. 
Recently, various Lindbladian QMEs were derived without resorting to the secular approximation
from the Redfield equation which does not guarantee the complete positivity. Here, we explicitly calculate, in a perturbative manner, the equilibrium steady states of these Lindbladian QMEs. We compare the results with the steady state of the Redfield equation 
obtained from an analytic continuation method, which coincides with the so-called mean force Gibbs (MFG) state.
The MFG state is obtained by integrating out the bath degrees of freedom for the Gibbs state of the total Hamiltonian. We explicitly show that the steady states of the Lindbladian QMEs are different from the MFG state. Our results indicate that manipulations of the Redfield equation needed to enforce complete positivity of a QME drives its steady state away from the MFG state. We also find that, in the high-temperature regime, both the steady states of the Lindbladian QMEs and MFG state reduce to the same Gibbs state of a system Hamiltonian under certain conditions.

\end{abstract}
%%%%%%%%%%%%%%%%%%%%%%%%%%%%%%%%%%%%%%%%%%%%%%%%%%%%%%%%%%%%%%%%%%%%%%%
%\pacs{}

%%%%%%%%%%%%%%%%%%%%%%%%%%%%%%%%%%%%%%%%%%%%%%%%%%%%%%%%%%%%%%%%%%%%%%%
 \maketitle

%%%%%%%%%%%%%%%%%%%%%%%%%%%%%%%%%%%%%%%%%%%%%%%%%%%%%%%%%%%%%%%%%%%%%%%
\section{Introduction}

Many interesting fundamental and practical physical problems arise in open quantum systems \cite{breuer, rivas} interacting with an environment acting as a thermal bath. One of the most widely used equations to study open quantum systems is the quantum master equation (QME), which describes the temporal evolution of the reduced density operator of the system. The reduced density operator is obtained by tracing over the variables describing the environment, for which an explicit description of dynamics is not usually feasible and often not of interest.

Various forms of QMEs can be derived depending on the underlying physical situation or assumptions. When a system is weakly coupled to the environment, the Born-Markov approximation can be applied~\cite{breuer,rivas}, from which the Redfield equation can be derived~\cite{redfield}. This master equation can also be obtained from the more formal Nakajima-Zwanzig time-convolutionless “projection operator perturbation expansion”~\cite{nakajima,zwanzig} via truncation at the second order of the interaction strength between the system and bath. However, the Redfield equation does not guarantee complete positivity of the density matrix, which may give rise to unphysical solutions. The most general form of the Markovian QME that preserves the positivity is known as the Lindblad equation~\cite{lindblad,gks}. A weak coupling limit~\cite{davies,spohn}, which amounts to the so-called secular approximation, is needed to obtain the QME in Lindblad form from the Redfield equation.

The secular approximation is usually valid in a quantum optics setting, where the energy level spacings are sufficiently large, but fails when there are nearly degenerate energy levels. Recently, many attempts have been made \cite{ule,becker,trush,david,potts,becker2} to derive  a completely positive QME in Lindblad form from the Redfield equation without resorting to secular approximation. These studies demonstrate that there is no specific way to derive a QME in Lindblad form: as noted in Ref.~\cite{ule}, there is a continuous family of Markovian QMEs within the same order of precision. This naturally raises the question of which QME is most appropriate to describe a given open quantum system, and the physical basis on which a given QME can be justified.

One way to examine the appropriateness of QMEs is to study their thermodynamic behavior~\cite{binder}, and especially their steady states. Here, we focus on the \emph{equilibrium} steady state of QMEs when the system is coupled with a single thermal bath. It is well known~\cite{breuer} that the Lindblad equation obtained from secular approximation results in the Gibbs state of the system Hamiltonian as a steady state. However, if we accept that the total system, consisting of the system of interest and the bath, thermalizes, for example by satisfying the condition required for the eigenstate thermalization hypothesis~\cite{sred,deutsch,Huse,Rigol}, the steady state of the reduced density operator will be the so-called mean force Gibbs (MFG) state~\cite{mfg,cresser,subasi,lobejko}, which is obtained by tracing the Gibbs state of the total Hamiltonian over the bath variables~\cite{Muller,Liu,Garrison}. Therefore, the Lindblad equation is not consistent with the thermalization property because it neglects the effect of coupling to the bath. In comparison, the Redfield equation correctly takes the MFG state, expanded up to the second order of coupling between the system and bath, as its steady state~\cite{thingna}. The perturbative evaluation of the steady state of the Redfield equation has some subtleties, since only the second-order coherence (off-diagonal) components can be determined self-consistently~\cite{mori,fleming,geva}. In general, correction to the population (diagonal) components in the steady state requires fourth-order expansion of the Redfield equation~\cite{mori,fleming,geva}. However, the analytic continuation method, which has been
developed in Ref.~\cite{thingna} and applied to various situations in Ref.~\cite{thingna2,xu}, provides a \emph{convenient} way to evaluate the second-order population within the Redfield equation; the result coincides with the second-order coherence and population of the MFG state~\cite{thingna}. 

In this respect, the steady states of the completely positive QMEs mentioned above, which are obtained without resorting to secular approximation, are of great interest. It was recently noted~\cite{tupkary} that any modification of the Redfield equation to achieve complete positivity inevitably changes the perturbative steady state from ~the MFG state. In this paper, we explicitly evaluate the steady states of these Lindbladian QMEs perturbatively. We focus on two prominent examples: the universal Lindblad equation (ULE)~\cite{ule,nathan} and truncated Lindblad equation (TLE)~\cite{becker}. We use the analytic continuation method of Ref.~\cite{thingna} to evaluate the second-order population, as well as the coherence. We find that the Gibbs state of the system Hamiltonian is not a steady state, but the effect of coupling to the bath is included in the solution (unlike the case of the secular approximation). We explicitly compare the steady states of these QMEs with the MFG state. We find that although the steady states of these Lindbladian equations capture the effect of coupling to the bath, they differ from the MFG state. This can be regarded as a tradeoff for Lindbladian QMEs derived from the Redfield equation. To ensure that the coupling effect persists, as in the MFG state, complete positivity must be sacrificed and vice versa. 

We also evaluate the steady states of QMEs in a high-temperature regime where the order of the inverse temperature is similar to the system-bath coupling strength. From this calculation, we find that the equilibrium steady states of various QMEs can become the Gibbs state of the system Hamiltonian up to the second order of coupling strength under certain conditions. Therefore, the steady state of the conventional Lindblad equation is also acceptable in this high-temperature regime, even when the system–bath coupling strength is weak but finite. 

This paper is organized as follows. In Section II, we describe how to calculate the steady state of the Redfield equation and MFG state perturbatively.  In Section III, we calculate the steady states of the ULE and TLE up to the second order of coupling strength.  We conclude the paper in Section IV.

%%%%%%%%%%%%%%%%%%%%%%%%%%%%%%%%%%%%%%%%%%%%%%%%%%%%%%%%%%%%%%%%%%%%%%%

\section{Redfield equation and its steady state}
In this section we review the calculation of perturbative steady states of the Redfield equation. 
We consider a quantum mechanical system interacting with the bath described by the Hamiltonians $H_{\rm S}$
and $H_{\rm B}$, respectively. The total Hamiltonian is written as
\begin{align}
H=H_{\rm S}+H_{\rm B}+H_{\rm I},
\label{total}
\end{align}
where the interaction between the system and the bath is governed by $H_{\rm I}$. 
We assume that the initial state is given by the product of the system and bath 
states as $\rho(0)\otimes\rho_{\rm B}$, where $\rho_{\rm B}$ describes the 
equilibrium state of the bath with the inverse temperature $\beta$. 
The state of the system at time $t$ is given by the reduced density operator ($\hbar=1$)
\begin{align}
\rho(t)=\mathrm{Tr}_{\rm B} (e^{-iHt}\rho(0)\otimes\rho_B e^{iHt}),
\end{align}
where the partial trace over the bath is performed. 

In general, a QME describes the time evolution of the
reduced density operator expressed as
\begin{align}
\frac{d\rho(t)}{dt}=\mathcal{L}[\rho(t)].
\label{qme}
\end{align} 
Various forms of the superoperator $\mathcal{L}$ can be derived especially in the regime where
the coupling between the system and the bath is weak. 
The Redfield equation can either be obtained
by applying the Born-Markov approximation or by using the time-convolutionless projector operator formalism \cite{breuer,rivas}. 
In the interaction picture, denoted by tilde as $\tilde{\rho}(t) = e^{i H_{\rm S} t} \rho(t) e^{-i H_{\rm S} t} $ and $\tilde{H}_{\rm I} (t) = e^{i H_{\rm S} t} H_{\rm I} e^{-i H_{\rm S} t} $, it is given by
\begin{align}
\frac{d\tilde{\rho}(t)}{dt}=-\int_0^t ds\; \mathrm{Tr}_{\rm B} [ \tilde{H}_{\rm I} (t),[\tilde{H}_I(t-s),\tilde{\rho}(t)\otimes\rho_{\rm B}]].
\label{redfield}
\end{align}
In some cases, a further approximation is made where the upper limit of the integral is taken to infinity. This is also referred to 
as the Redfield equation in literature.

We consider the case where the interaction
Hamiltonian takes the form
\begin{align}
H_{\rm I}=\epsilon \sum_\alpha A_\alpha\otimes B_\alpha ,
\label{interaction}
\end{align}
where $\epsilon$ measures the strength of the coupling between the system and bath
and $A_\alpha$ and $B_\alpha$ are (hermitian) operators acting on the system and bath spaces, respectively.
It is convenient to use the expansion using the Bohr frequency as
\begin{align}
A_\alpha(\omega)\equiv  \sum_{\substack{m,n\\ E_n-E_m=\omega}} \Pi_m A_\alpha \Pi_n ,
\end{align}
where $\Pi_m=\vert m\rangle\langle m\vert $ is the projection operator onto the eigenstate $\vert m\rangle$ of $H_S$
with eigenvalue $E_m$. Note that $A_\alpha^\dag(\omega)=A_\alpha(-\omega)$.
Then we can rewrite Eq.~(\ref{redfield}) as
\begin{align}
\frac{d\tilde{\rho}(t)}{dt}=-\epsilon^2 \sum_{\alpha,\beta}\sum_{\omega,\omega^\prime}&\Big(e^{i(\omega^\prime-\omega)t}
\Gamma^t_{\alpha\beta}(\omega)[A^\dag_\alpha(\omega^\prime),A_\beta(\omega)\tilde{\rho}(t)] \nonumber \\
&+\mathrm{h.c.}\Big), 
\label{red_int}
\end{align}
where $\Gamma^t_{\alpha\beta}(\omega)$ is defined as
\begin{align}
\Gamma^t_{\alpha\beta}(\omega) \equiv \int_0^t ds\; e^{i\omega s}C_{\alpha\beta}(s).
\end{align}
Here, the bath correlation function $C_{\alpha\beta}(t)$ is given by 
\begin{align}
C_{\alpha\beta}(t)\equiv\mathrm{Tr}_B [\tilde{B}_\alpha(t) B_\beta\rho_B]
\end{align}
with $\tilde{B}_\alpha(t)\equiv e^{iH_B t}B_\alpha e^{-iH_B t}$. Note that $C_{\alpha\beta}^*(t)=C_{\beta\alpha}(-t)$ due to the time translational invariance.

In the Schr\"odinger picture, the superoperator in Eq.~(\ref{qme}) for the Redfield equation takes the form,
$\mathcal{L}_{\rm Red}=\mathcal{L}^{(0)}+\epsilon^2\mathcal{L}^{(2)}_{\rm Red}$, where
$\mathcal{L}^{(0)}[\rho(t)]=-i[H_S,\rho(t)]$ and
\begin{align}
\mathcal{L}^{(2)}_{\rm Red}[\rho(t)]=-\sum_{\alpha,\beta}\sum_{\omega, \omega^\prime } &\Big( \Gamma^t_{\alpha\beta}(\omega)
 [A_\alpha^\dagger (\omega^\prime) ,A_\beta(\omega)\rho(t)] \nonumber \\
&+\mathrm{h.c.}\Big),
 \label{L2_red1}
\end{align}
or equivalently
\begin{align}
\mathcal{L}^{(2)}_{\rm Red}[\rho(t)]=-\sum_{\alpha,\beta}\int_0^t ds\; &\Big( C_{\alpha\beta}(s)
 [A_\alpha,\tilde{A}_\beta(-s)\rho(t)] \nonumber \\
 &+\mathrm{h.c.}\Big),
 \label{L2_red}
\end{align} 
with $\tilde{A}_\alpha(t)\equiv e^{iH_S t}A_\alpha e^{-iH_S t}$.

%##############################

In order to find the steady state of the Redfield equation, we first decompose
Eq.~(\ref{L2_red1}) (after taking $t\to \infty$) into $\mathcal{L}^{(2)}_{\mathrm{Red}}=
\mathcal{L}^{(S)}_{\mathrm{Red}}+\mathcal{L}^{(\gamma)}_{\mathrm{Red}}$, where
\begin{align}
\mathcal{L}^{(S)}_{\mathrm{Red}}[\rho(t)]=&-i \sum_{\alpha,\beta}\sum_{\omega,\omega^\prime}S_{\alpha\beta}(\omega)
\Big[ -A_\beta(\omega)\rho(t) A^\dag_\alpha(\omega^\prime) \nonumber \\
&+A_\beta(\omega^\prime)\rho(t) A^\dag_\alpha(\omega) +A^\dag_\alpha(\omega^\prime) A_\beta(\omega) \rho(t) \nonumber \\
&-\rho(t) A^\dag_\alpha(\omega) A_\beta(\omega^\prime)\Big],~~{\rm and} \label{LS} \\
\mathcal{L}^{(\gamma)}_{\mathrm{Red}}[\rho(t)]=&\frac 1 2 \sum_{\alpha,\beta}\sum_{\omega,\omega^\prime}\gamma_{\alpha\beta}(\omega)
\Big[ A_\beta(\omega)\rho(t) A^\dag_\alpha(\omega^\prime) \nonumber \\
&+A_\beta(\omega^\prime)\rho(t) A^\dag_\alpha(\omega) -A^\dag_\alpha(\omega^\prime) A_\beta(\omega) \rho(t) \nonumber \\
&-\rho(t) A^\dag_\alpha(\omega) A_\beta(\omega^\prime)\Big]. \label{Lg}
\end{align}
Here we have introduced $\gamma_{\alpha\beta}(\omega)\equiv\Gamma^\infty_{\alpha\beta}(\omega)+
\Gamma^{\infty\ast}_{\beta\alpha}(\omega)=\int_{-\infty}^\infty dt  e^{i\omega t} C_{\alpha\beta}(t)=\gamma^*_{\beta\alpha}(\omega)$ and
$S_{\alpha\beta}(\omega)\equiv (\Gamma^\infty_{\alpha\beta}(\omega)-
\Gamma^{\infty\ast}_{\beta\alpha}(\omega))/(2i)$. These two are related to each other as
\begin{align}
	S_{\alpha\beta}(\omega)=-\mathcal{P}\int^\infty_{-\infty}\frac{d\tilde{\omega}}{2\pi}\frac 1{\tilde{\omega}} \gamma_{\alpha\beta}(\tilde{\omega}+\omega),
	\label{Sab}
\end{align}
where $\mathcal{P}$ denotes the Cauchy principal value.
As noted in Introduction, the system Gibbs state
$\rho_\mathrm{G}=e^{-\beta H_\mathrm{S}}/Z_{\rm S} $ 
of the system Hamiltonian with $Z_{\rm S}=\mathrm{Tr_S}e^{-\beta H_\mathrm{S}}$ is not a steady state of the Redfield equation.
It is, however, interesting to note that we still have
$\mathcal{L}^{(\gamma)}_{\mathrm{Red}}[\rho_{\rm G}]=0$. This can be proved by using
$\rho_\textrm{G} A_\alpha (\omega) = e^{\beta \omega} A_\alpha(\omega) \rho_\textrm{G}$, $\rho_\textrm{G} A_\alpha^\dagger(\omega) = e^{-\beta \omega} A_\alpha^\dagger (\omega) \rho_\textrm{G}$, and the Kubo-Martin-Schwinger (KMS) condition~\cite{kms,kms1} $\gamma_{\alpha\delta}(-\omega)=
e^{-\beta\omega}\gamma_{\delta\alpha}(\omega)$
satisfied by the bath in equilibrium at temperature $\beta^{-1}$.
It is the particular form of $\mathcal{L}^{(S)}_{\mathrm{Red}}$ that 
prevents $\rho_{\rm G}$ from being the steady state of the Redfield equation as we can easily see that
$\mathcal{L}^{(S)}_{\mathrm{Red}}[\rho_{\rm G}]\neq 0$ in general. 

\subsection{Perturbative steady-state solution of Redfield equation}

One can calculate the steady of the Redfield equation in a perturbative way. 
In this subsection, we review the method developed in
Ref.~\cite{thingna} for its calculation. In the next section, we will apply this method to 
the completely positive master equations to calculate their perturbative steady states.
By regarding the Redfield superoperator as an $O(\epsilon^2)$ expansion of the full expression and by
expanding the steady state as
$\rho^{\rm st}=\rho^{(0)}+\epsilon^2 \rho^{(2)}+O(\epsilon^4)$, we have steady-state equations, 
$\mathcal{L}^{(0)}[\rho^{(0)}]=0$ and $\mathcal{L}^{(0)}[\rho^{(2)}]+\mathcal{L}^{(2)}_{\rm Red}[\rho^{(0)}]=0$.
In terms of the matrix elements, $\rho_{nm}=\langle n\vert \rho \vert m\rangle$, with respect to the eigenstates of $H_\mathrm{S}$, the former leads to
\begin{align}
\rho^{(0)}_{nm}=0,~~~~\mathrm{for}~~n\neq m,
\label{st_eq0}
\end{align}
and the latter becomes
\begin{align}
-i\Delta_{nm}\rho^{(2)}_{nm}+\left( \mathcal{L}^{(2)}_{\rm Red}[\rho^{(0)}]\right)_{nm}=0,
\label{st_eq2}
\end{align}
where $\Delta_{nm}=E_n-E_m$. 

The diagonal part of Eq.~(\ref{st_eq2}),
\begin{align}
\left( \mathcal{L}^{(2)}_{\rm Red}[\rho^{(0)}]\right)_{nn}=0
\label{eq_dia}
\end{align}
determines $\rho^{(0)}$. 
We can explicitly show from Eq.~(\ref{LS}) that $\left( \mathcal{L}^{(S)}_{\mathrm{Red}}[\rho^{(0)}]\right)_{nn}$ vanishes
from the fact that $\rho^{(0)}$ is diagonal.
Therefore, the condition, Eq.~(\ref{eq_dia}) reduces to
\begin{align}
0&=\left( \mathcal{L}^{(\gamma)}_{\mathrm{Red}}[\rho^{(0)}]\right)_{nn} \\
&=
\sum_{\alpha,\beta}\sum_j \left(\gamma_{\alpha\beta}(\Delta_{jn})\rho^{(0)}_{jj}-\gamma_{\beta\alpha}(\Delta_{nj})\rho^{(0)}_{nn}\right)
(A_\beta)_{nj}(A_\alpha)_{jn} . \nonumber
\end{align}
The KMS condition for $\gamma_{\alpha\beta}$ implies that $\rho^{(0)}=\rho_{\rm G}$. 
The off-diagonal part of Eq.~(\ref{st_eq2}) gives the off-diagonal element of the $O(\epsilon^2)$-correction to the 
reduced density matrix. We have for $m\neq n$
\begin{align}
\rho^{(2)}_{nm}=\frac{1}{i\Delta_{nm}}\left( \mathcal{L}^{(S)}_{\mathrm{Red}}[\rho^{(0)}]\right)_{nm},
\end{align}
where we have used the fact that $\left( \mathcal{L}^{(\gamma)}_{\mathrm{Red}}[\rho_{\rm G}]\right)_{nm}$
vanishes as mentioned above. From Eq.~(\ref{LS}), we have for $n\neq m$,
\begin{align}
\rho^{(2)}_{nm}&=\frac{1}{\Delta_{nm}}
\sum_{\alpha,\beta, j}\Big[ \{ S_{\beta\alpha}(\Delta_{jn})-S_{\beta\alpha}(\Delta_{jm})\} \rho^{(0)}_{jj} \label{rho2nm} \\
&+\left\{S_{\alpha\beta}(\Delta_{nj})\rho^{(0)}_{nn} -S_{\alpha\beta}(\Delta_{mj})\rho^{(0)}_{mm} \right\} \Big]
(A_{\alpha})_{nj}(A_\beta)_{jm}. \nonumber
\end{align}
Up to $O(\epsilon^2)$, the diagonal part $\rho^{(2)}_{nn}$ is not determined in this perturbative scheme. In Ref.~\cite{thingna},
$\rho^{(2)}_{nn}$ was obtained through
an analytic continuation which takes $m\to n$ in Eq.~(\ref{rho2nm}). We note that this analytic continuation method
is not related to any analytic property of the spectral function. Rather it is a simple limiting procedure. Since $\Delta_{nm}$ as well as the numerator in
Eq.~(\ref{rho2nm}) vanishes in the limit $m\to n$, the analytic continuation method just involves taking the
derivative $S_{\alpha\beta}^\prime(\omega)=dS_{\alpha\beta}(\omega)/d\omega$. In our notation, we obtain
$\bar{\rho}^{(2)}_{nn}\equiv\lim_{m\to n}\rho^{(2)}_{mn}$ as
\begin{align}
\bar{\rho}^{(2)}_{nn}=\sum_{\alpha,\mu, j}&\Big[ -S^\prime_{\mu\alpha}(\Delta_{jn})\rho^{(0)}_{jj}+S^\prime_{\alpha\mu}(\Delta_{nj})\rho^{(0)}_{nn} \nonumber \\
&-\beta S_{\alpha\mu}(\Delta_{nj})\rho^{(0)}_{nn}\Big] (A_\alpha)_{nj}(A_\mu)_{jn}.
\label{rho2nn}
\end{align}
Note that the zero frequency part of $S^\prime(\omega)$ does not appear in the above equation since
the $j=n$ terms are all cancelled out.
The normalization $\mathrm{Tr}[\rho^{(0)}+\epsilon^2\rho^{(2)}]=1$ requires that the correction to population is actually
$\rho^{(2)}_{nn}=\bar{\rho}^{(2)}_{nn}-\rho^{(0)}_{nn} \sum_k \bar{\rho}^{(2)}_{kk}$.

%##############################

\subsection{Mean Force Gibbs State}
\label{subsec:mfg}
Since we expect that a quantum system coupled to a thermal bath will equilibriate, the steady state of
a QME must be described by the MFG state
\begin{align}
	\rho_{\rm mG} \equiv  \frac{{\rm Tr}_{\rm B} (e^{-\beta H})}{{\rm Tr}_{\rm SB} (e^{-\beta H} )} . \label{eq:rG}
\end{align}
In the weak coupling regime,
%$\epsilon \ll \epsilon_0$, where $\epsilon_0$ is the typical energy scale of the system, 
$\rho_{\rm mG}$ can be expanded up to the second order in $\epsilon$ as explained in Ref.~\cite{thingna,cresser}. 
The detailed calculation is also presented in Appendix \ref{sec:app_CPT} for self-containedness. The result is
\begin{align}
	\rho_{\rm mG} = \rho_{\rm G} + \epsilon^2 \rho^{(2)}_{\rm mG} + O(\epsilon^3). \label{eq:rho_rG_expansion}
\end{align}
The second-order correction is given by
\begin{align}
	\rho^{(2)}_{\rm mG}& = \frac{\mathcal{D}}{Z_{\rm S}} - \frac{ {\rm Tr}_{\rm S} (\mathcal{D})}{Z_{\rm S}^2} e^{-\beta H_{\rm S}} 
\end{align}
and 
\begin{align}
	\mathcal{D} = &\int_0^\beta d \lambda_1 \int_0^{\lambda_1} d \lambda_2~ e^{-\beta H_{\rm S}} \nonumber \\
	&\times \sum_{\alpha,\gamma} \tilde{A}_\alpha (-i \lambda_1) \tilde{A}_\gamma (-i \lambda_2) C_{\alpha\gamma}(-i(\lambda_{1}-\lambda_2)) \label{eq:D}
\end{align}
with $\tilde{A}_\alpha (u) \equiv e^{i u H_{\rm S}} A_\alpha e^{-i u H_{\rm S}} $. 
We can show \cite{thingna} that the second-order correction  $\rho^{(2)}$ of the steady state of the Redfield equation is exactly equal to $\rho^{(2)}_{\rm mG}$. We include the detailed calculations in Appendix \ref{sec:app_CPT}.  
It amounts to showing that $\mathcal{D}_{nm}/Z_{\rm S}=\rho^{(2)}_{nm}$ in Eq.~(\ref{rho2nm}) for $n\neq m$ 
and  $\mathcal{D}_{nn}/Z_{\rm S}=\bar{\rho}^{(2)}_{nn}$ in Eq.~(\ref{rho2nn}), where the matrix elements $\mathcal{D}_{nm} = \langle n| \mathcal{D} |m \rangle $. Compared to the other derivations in Refs.~\cite{thingna} and \cite{cresser},
we do not use a particular form of a bosonic bath, but use general analytic properties of the bath
correlation functions. Our method applies to any bath whose
correlation function, when integrated over a contour (shown in Fig.~\ref{fig:mfg} in Appendix \ref{sec:app_CPT}) on the complex plane, vanishes.
The calculations in Refs.~\cite{thingna} and \cite{cresser} are special cases.

Now we consider both the weak coupling and high temperature regime, where we can 
expand  Eq.~\eqref{eq:rho_rG_expansion} further in power series of $\beta$. In particular we are interested in the regime
where $\beta$ is of the same order of the coupling $\epsilon$.
%satisfying the condition $\epsilon/\epsilon_0 \sim \beta \epsilon_0 \ll 1$. 
%Then, further expansion of Eq.~\eqref{eq:rho_rG_expansion} can be done in terms of $\beta$. 
As most QMEs describe dynamics of the subsystem up to the second order in $\epsilon$, we study the small-$\epsilon$ and high-temperature expansion up to the order in $\epsilon^{2-n} \beta^n$. The expansion of $\rho_{\rm G}$ in Eq.~\eqref{eq:rho_rG_expansion} is straightforward, thus, here we focus on the expansion of $\rho_{\rm mG}^{(2)}$. As the integration range $\int_0^\beta d \lambda_1 \int_0^{\lambda_1} d \lambda_2$ of $\mathcal{D}$ in Eq.~\eqref{eq:D} gives $\beta^2$ order, $\rho_{\rm mG}^{(2)}$ as a whole is $O(\beta^2 C_{\alpha\gamma}/Z_{\rm S})$. Therefore, when $O(C_{\alpha\gamma}/Z_{\rm S}) < O (\beta^{-2}) $, the MFG state simply becomes $\rho_{\rm G}$, that is 
\begin{align}
	\rho_{\rm mG} = \rho_{\rm G} + &\textrm{ higher order than } \epsilon^2.  \label{eq:rho_rG_highT}
\end{align}
Therefore, the system Gibbs state $\rho_{\rm G}$ is regarded as a steady state of a system weakly coupled to a bath in the high temperature regime when $O(C_{\alpha\gamma}/Z_{\rm S}) < O (\beta^{-2})$.

%%%%%%%%%%%%%%%%%%%%%%%%%%%%%%%%%%%%%%%%%%%%%%%%%%%%%%%%%%%%%%%%%%%%%%%
\section{Steady states of Lindbladian equations derived from the Redfield equation}
In this section, we present the perturbative steady states of the completely positive quantum master equations
which are not based on the secular approximation. For a comparison purpose, we first collect the results for the
Lindblad equation obtained via the secular approximation.

\subsection{Lindblad equation from the secular approximation}

When the system energy level spacing is large compared to 
the inverse of the relaxation time scale, the secular approximation can be applied. It amounts to taking only the diagonal part ($\omega=\omega'$)  
in the double sums in Eqs.~(\ref{LS}) and (\ref{Lg}). After taking the upper limit of the integral in Eq.~(\ref{redfield}) to infinity,
we have the superoperator for the secular approximation as
$\mathcal{L}_{\rm sec}=\mathcal{L}^{(0)}+\epsilon^2[\mathcal{L}^{(S)}_{\rm sec}+\mathcal{L}^{(\gamma)}_{\rm sec}]$.
It is straightforward to see that
\begin{align}
	\mathcal{L}^{(S)}_{\rm sec}[\rho(t)]=-i[H_{\rm LS},\rho(t)]
	\label{sec_l2}
\end{align}
where
\begin{align}
	H_{\rm LS}=\sum_{\alpha,\beta}\sum_\omega S_{\alpha\beta}(\omega)A^\dag_\alpha(\omega)A_\beta(\omega)
	\label{Lamb}
\end{align}
and
\begin{align}
	\mathcal{L}^{(\gamma)}_{\rm sec}[\rho(t)]=\sum_{\alpha,\beta}\sum_\omega& \gamma_{\alpha\beta}(\omega) \Big( 
	A_\beta (\omega)\rho(t)A^\dag_\alpha (\omega) \nonumber \\
	&-\frac 1 2 \left\{ A^\dag_\alpha(\omega)A_\beta(\omega),
	\rho(t)\right\}\Big).
	\label{diss}
\end{align}
Since $\gamma_{\alpha\beta}(\omega)$ is positive semidefinite, Eq.~(\ref{diss}) is in the Lindblad form
guaranteeing the positivity of the reduced dynamics.

It is well known that $\rho_{\rm G}$ is a steady state for this equation \cite{breuer}. 
We have already established that $\mathcal{L}^{(\gamma)}_{\rm Red}[\rho_{\rm G}]=0$. It follows in a similar manner that
$\mathcal{L}^{(\gamma)}_{\rm sec}[\rho_{\rm G}]=0$.
Unlike the Redfield equation, $\mathcal{L}^{(S)}_{\rm sec}$ now takes the form of the unitary evolution as shown in 
Eq.~(\ref{sec_l2}), which gives $\mathcal{L}^{(S)}_{\rm sec}[\rho_{\rm G}]=0$. This can be easily checked by using the relation $A^\dag_\alpha(\omega)A_\beta(\omega) \rho_{\rm G} = \rho_{\rm G} A^\dag_\alpha(\omega)A_\beta(\omega)$.
Although the Lindblad equation obtained from the secular approximation is of $O(\epsilon^2)$,
it does not give $O(\epsilon^2)$ corrections to $\rho_{\rm G}$ in the steady state. Therefore, as far as the steady state is concerned,
the secular approximation is valid in the $\epsilon\to 0$ limit \cite{mfg}.

\subsection{Steady state of the universal Lindblad equation}
\label{subsec:ule}

In Ref.~\cite{ule}, a QME called the ULE has been derived, 
which preserves the positivity of the reduced dynamics.
The derivation does not rely on the secular approximation and has the same level of accuracy as the Redfield equation.  
The main ingredient of the approximation is the identification
of a small parameter given in terms of the properties of the bath valid in the weak-coupling limit.
This results in 
the jump operator for the Lindblad form given by the square root of the spectral function \cite{ule}.
In this section, we calculate the perturbative steady state of the ULE.
We focus on the case of the time-independent system Hamiltonian, in which case the ULE takes the form in the Schr\"odinger picture 
$\partial_t \rho(t)=\mathcal{L}^{(0)}[\rho(t)]+\epsilon^2 \mathcal{L}^{(2)}_{\rm ULE}[\rho(t)]$ \cite{ule}, where
the $O(\epsilon^2)$ superoperator has two components: $\mathcal{L}^{(2)}_{\rm ULE}=\mathcal{L}^{(a)}_{\rm ULE}
+\mathcal{L}^{(b)}_{\rm ULE}$. The first one takes the form similar to the Lamb-shift term as
$\mathcal{L}^{(a)}_{\rm ULE}[\rho(t)]=-i[\Lambda,\rho(t)]$ with
\begin{align}
\Lambda &= \sum_{\alpha,\beta} \sum_{\omega,\omega^\prime} f_{\alpha\beta} (\omega,\omega^\prime) A_\alpha(\omega) A_\beta(\omega^\prime),
\label{eq:Lambda}
\end{align}
where $f_{\alpha\beta}(\omega,\omega^\prime)$ is defined in Eq.~\eqref{eq:f}. 
The second one acts as a dissipator in the Lindblad form given as
\begin{align}
\mathcal{L}^{(b)}_{\rm ULE}[\rho(t)]=
\sum_\alpha [ L_\alpha \rho(t) L_\alpha^\dagger - \frac{1}{2} \{ L_\alpha^\dagger L_\alpha, \rho(t)  \}] 
\label{ule_l2}
\end{align}
with the jump operator
\begin{align}
L_\alpha &=   \sum_{\beta}\sum_\omega g_{\alpha\beta}(\omega) A_\beta(\omega).
\label{eq:L}
\end{align}
Here $g_{\alpha\beta}(\omega)$ is a matrix square root of the Fourier transform $\gamma_{\alpha\beta}(\omega)$ 
of the bath correlation function given as~\cite{g_convention}
\begin{align}
\gamma_{\alpha\beta}(\omega)=\sum_\mu g_{\alpha\mu}(\omega)g_{\mu\beta}(\omega).
\label{def:g}
\end{align}
Since $\gamma_{\alpha\beta}(\omega)$ is positive semidefinite, $g_{\alpha\beta}(\omega)$ is hermitian and also positive semidefinite.
Now the function  $f_{\alpha\beta}(\omega,\omega^\prime)$ is given by
\begin{align}
f_{\alpha\beta}(\omega,\omega^\prime) \equiv  - \mathcal{P} \sum_\mu \int_{-\infty}^\infty \frac{d \tilde{\omega}}{2\pi}
\frac{1}{\tilde{\omega}} g_{\alpha\mu}(\tilde{\omega}-\omega)g_{\mu\beta}(\tilde{\omega}+\omega^\prime).
\label{eq:f}
\end{align}
A comparison of the above equations with Eqs.~(\ref{Sab}), (\ref{sec_l2}), (\ref{Lamb}), and (\ref{diss}) for the secular approximation
reveals that the ULE roughly corresponds to taking
the square root of the Fourier transform of the bath correlation function as defined in Eq.~(\ref{def:g}) and to distributing it to the new jump operators.

We can check whether the system Gibbs state $\rho_\textrm{G}$
is a steady state solution of the ULE. Using Eqs.~(\ref{eq:Lambda}) and (\ref{eq:L}), it is in fact straightforward to
verify that both $\mathcal{L}^{(a)}_{\rm ULE}[\rho_{\rm G}]$ and $\mathcal{L}^{(b)}_{\rm ULE}[\rho_{\rm G}]$ in general
do not vanish. The detailed calculations are in Appendix \ref{sec:app1}.

We now apply the perturbation scheme used for the Redfield equation to the ULE to find its stationary state. 
We again write the steady state as $\rho^{\rm st}=\rho^{(0)}+\epsilon^2 \rho^{(2)}+O(\epsilon^4)$.
Since $\mathcal{L}^{(0)}$ takes the same form as the Redfield equation, the off-diagonal matrix elements of $\rho^{(0)}$ vanish. 
The diagonal 
matrix element of $\rho^{(0)}$ is determined by
\begin{align}
\left( \mathcal{L}^{(2)}_{\rm ULE}[\rho^{(0)}]\right)_{nn}=0.
\label{eq_dia_ule}
\end{align}
We can show that $\mathcal{L}^{(a)}_{\rm ULE}[\rho^{(0)}]_{nn}$ identically vanishes (see Appendix \ref{sec:app2}
for details). Now the condition $(\mathcal{L}^{(b)}_{\rm ULE}[\rho^{(0)}])_{nn}=0$ yields
\begin{align}
\sum_{\alpha,\beta,\gamma}&\sum_{k}\big[ g_{\alpha\beta}(\Delta_{kn})
g_{\gamma\alpha}(\Delta_{kn})\rho^{(0)}_{kk} \nonumber \\
&-g_{\beta\alpha}(\Delta_{nk})g_{\alpha\gamma}(\Delta_{nk})\rho^{(0)}_{nn}\big](A_\beta)_{nk}(A_{\gamma})_{kn}=0.
\label{cond1_ule}
\end{align}
Note that the KMS condition for $\gamma_{\alpha\delta}(\omega)$ 
translates into $g_{\alpha\delta}(-\omega)=e^{-\beta\omega/2}g_{\delta\alpha}(\omega)$. By applying this, we immediately find that 
$\rho^{(0)}=\rho_{\rm G}$ satisfies the above condition. As in the Redfield equation, the zeroth order of the
steady state is again given by $\rho_{\rm G}$.

The off-diagonal elements of the $O(\epsilon^2)$ part of the steady state is obtained through the condition corresponding to Eq.~(\ref{st_eq2}) as
\begin{align}
\rho^{(2)}_{mn}=\frac{1}{i\Delta_{nm}}\left( \mathcal{L}^{(2)}_{\mathrm{ULE}}[\rho^{(0)}]\right)_{nm} \label{eq:ULE_2nd}
\end{align}
for $n\neq m$. We have evaluated Eq.~\eqref{eq:ULE_2nd} for $n\neq m$ in Appendix \ref{sec:app2} as 
\begin{widetext}
\begin{align}
\rho^{(2)}_{nm}=&\frac{1}{i\Delta_{nm}}\Bigg[ -i \sum_{\alpha,\beta}\sum_{k}
f_{\alpha\beta}(\Delta_{kn},\Delta_{mk}) (A_\alpha)_{nk}(A_\beta)_{km} \left(\rho^{(0)}_{mm}-\rho^{(0)}_{nn}\right) \nonumber \\
&+ \sum_{\alpha,\beta,\gamma}\sum_{k} 
(A_\beta)_{nk}(A_\gamma)_{km} \Big\{
g_{\alpha\beta}(\Delta_{kn})g_{\gamma\alpha}(\Delta_{km}) \rho^{(0)}_{kk}
% \nonumber \\
-\frac 1 2 g_{\beta\alpha}(\Delta_{nk})g_{\alpha\gamma}(\Delta_{mk})\left(\rho^{(0)}_{mm}+\rho^{(0)}_{nn}\right) \Big\}\Bigg] .
\label{rho2nm_ule}
\end{align}
\end{widetext}
We note that this is manifestly different from the perturbative steady state, Eq.~(\ref{rho2nm}) of the Redfield equation. The latter
involves only the Lamb shift part for $\rho^{(2)}_{nm}$ for $n\neq m$. On the other hand, for the ULE, the $O(\epsilon^2)$ part
of the steady state also depends on the dissipator. We therefore conclude that the steady state of the ULE 
differs from both $\rho_{\rm G}$ and $\rho_{\rm mG}$, which are the steady states of the 
the QME with the secular approximation and the Redfield equation, respectively. 

The diagonal part, $\rho^{(2)}_{nn}$ can be obtained by the analytic continuation method used in Ref.~\cite{thingna}. We take
the limit $E_m\to E_n$ or $\theta\to 0$, where $E_m=E_n -\theta$. We first look at the dissipator part in the second term on
the right hand side of Eq.~(\ref{rho2nm_ule}). Using the KMS condition for $g_{\alpha\beta}(\omega)$ 
and the fact that $\rho^{(0)}_{nn}=(\rho_{\rm G})_{nn}\sim e^{-\beta E_n}$, we can show that the
quantity inside the curly bracket is proportional to 
\begin{align}
e^{-\beta(E_n+E_m)/2}-\frac 1 2 (e^{-\beta E_m}+e^{-\beta E_n}). \label{eq:ule_afterKMS}
\end{align}
In the limit $\theta\to 0$, this quantity is of $O(\theta^2)$.
Therefore the second term on the right hand side of Eq.~(\ref{rho2nm_ule}) does not contribute to 
$\rho^{(2)}_{nn}$. Taking the limit $\theta\to 0$ in the first term, we have
\begin{align}
\bar{\rho}^{(2)}_{nn}= -\beta \sum_{\alpha,\delta ,k}
f_{\alpha\delta}(\Delta_{kn},\Delta_{nk}) (A_\alpha)_{nk}(A_\delta)_{kn} \rho^{(0)}_{nn}.
\label{rho2nn_ule}
\end{align}
Unlike the off-diagonal counterparts, the diagonal part of $\rho^{(2)}$ depends only on the Lamb shift-like term. 
From Eqs.~(\ref{Sab}), (\ref{def:g}), and (\ref{eq:f}), we note that $f_{\alpha\delta}(\Delta_{kn},\Delta_{nk})
=S_{\alpha\delta}(\Delta_{nk})$. Therefore, Eq.~(\ref{rho2nn_ule}) is equal to only one term among three terms in 
the corresponding expression Eq.~(\ref{rho2nn}) for the Redfield equation. 
As in the Redfield equation, the second-order correction to the population is given by
$\rho^{(2)}_{nn}=\bar{\rho}^{(2)}_{nn}-\rho^{(0)}_{nn} \sum_k \bar{\rho}^{(2)}_{kk}$.

Now we discuss the behavior of the steady state of ULE in the weak coupling and the high temperature regime
as done in Sec.~\ref{subsec:mfg}.
% $\beta \epsilon_0 \sim \epsilon / \epsilon_0 \ll 1$. 
 We first consider the population term. In this regime, the orders of $\rho_{nn}^{(2)}$ is $O(\beta f_{\alpha \beta} /Z_{\rm S})$ from Eq.~\eqref{rho2nn_ule}. Therefore, when $O(f_{\alpha \beta} /Z_{\rm S}) < O(\beta^{-1})$, the population of the ULE steady state approaches that of the system Gibbs state. For the coherence term, the first summation part of the right hand side of Eq.~\eqref{rho2nm_ule} gives the order of $O(\beta f_{\alpha \beta}/Z_{\rm S})$. The order of the second summation part of the right hand side of Eq.~\eqref{rho2nm_ule} can be estimated as $O(\beta^2 g_{\alpha \beta}g_{\gamma \alpha}/Z_{\rm S})$ by using Eq.~\eqref{eq:ule_afterKMS}. Therefore, when $O(f_{\alpha \beta}/Z_{\rm S}) < O(\beta^{-1})$ and  $O(g_{\alpha \beta}g_{\gamma \alpha}/Z_{\rm S}) <O(\beta^{-2})$, the coherence of the ULE becomes negligible up to the order of $\epsilon^2$, which demonstrates that the steady state
of the ULE approaches the system Gibbs state $\rho_{\rm G}$ in this high temperature regime.

%##############################
\subsubsection{Example 1: Steady state of spin chain}

To confirm the validity of the analytic continuation method, we re-examine the simulation done in Section V A of Ref.~\cite{ule} for the relaxation process of the magnetization of a $N$ spin-chain system. The Hamiltonian of the spin chain is
\begin{equation}
	H_{\rm S} = - B_z \sum_{i=1}^N \sigma_i^z - \eta \sum_{i=1}^{N-1} \bm \sigma_i \cdot \bm \sigma_{i+1}, 
\end{equation}
where $B_z$ is a magnetic field strength, $\eta$ is a coupling strength between neighboring spins, and $\bm \sigma_i = (\sigma_i^x, \sigma_i^y, \sigma_i^z)$, where $\sigma_i^\mu$ denotes the $\mu$ component of the spin operator for the $i$th spin. One bath is coupled to $i=1$ spin through the $x$-component operator $S_1^x$ with a coupling strength $\epsilon$. The bath is in equilibrium with the Ohmic spectral density $J(\omega)$ of the bath correlation function as
\begin{equation}
	J(\omega) = \frac{2\pi}{\omega_0} \frac{\omega e^{-\frac{\omega^2}{2 \Lambda^2}}}{1-e^{-\beta \omega}}	,
\end{equation}
where $\omega_0$ is an energy scale, $\Lambda$ is a ultraviolet energy cutoff, and $\beta$ is the inverse temperature of the bath.

\begin{figure}
	\resizebox{0.90\columnwidth}{!}{\includegraphics{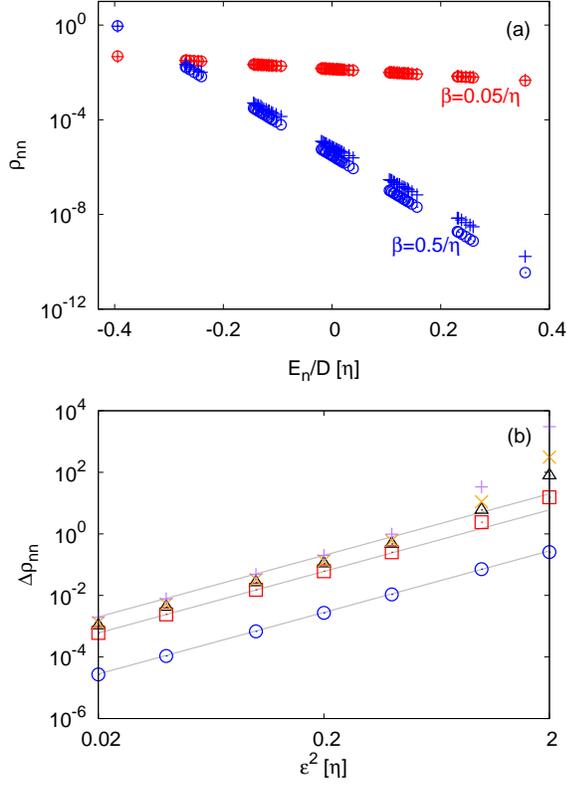}}
	\caption{ (a) Plot of the population $\rho_{nn}$ as a function of normalized energy eigenvalues of the system Hamiltonian with the normalization factor $D = 64$. $\bigcirc$ and $+$ denote $(\rho_{\rm G})_{nn}$ and $\rho_{nn}^{\rm ss}$, respectively. The blue and the red dots represent data for $\beta = 0.5/\eta$ and $\beta = 0.05/\eta$, respectively. The same plot for $\beta = 0.5 / \eta$ is presented in Fig.~1(b) of  Ref.~\cite{JSLeeComment}. 
	(b) Plot for $\Delta \rho_{nn}$ as a function of $\epsilon^2$. Gray solid lines denote $\epsilon^4$ curves as a guide for eye.  $\bigcirc$, $\square$, $\triangle$, $\times$, and $+$ are data for various $n$ associated with eigenenergy of the spin chain $E_n=-25.25$, $-8.75$, $0.25$, $7.25$, and $22.25$, respectively. In this plot $\epsilon^2$ has a dimension of $\eta$. }
	\label{fig1}
\end{figure}

We use the same parameter set as used in Ref.~\cite{ule}: $B_z = 8 \eta$, $\Lambda = 100 \eta$, $\omega_0 = 2\eta$, and $\beta = 0.5/ \eta$ for various $\epsilon^2$ within the range $ 0.02 \eta \leq  \epsilon^2 \leq 2 \eta$. The system is initially in the state with all spins aligned against $B_z$. As carried out in Ref.~\cite{ule}, we ignore the Lamb shift term. With these conditions, we numerically obtain the steady-state density matrix $\rho^{\rm ss}$ by directly solving the ULE for the spin chain with length $N = 6$. As expected, the obtained steady state deviates from the system Gibbs state of the system Hamiltonian as shown in Fig.~\ref{fig1}(a). Since the Lamb shift term is ignored, $\rho_{nn}^{(2)}$ term should vanish according to the analytic continuation method. Therefore, the next order of the diagonal term of the steady-state reduced density matrix for this ULE is $O(\epsilon^4)$, i.e., $\rho_{ nn}^{\rm ss} = (\rho_{\rm G})_{nn} + O(\epsilon^4)$. To verify this numerically, we make a plot of $\Delta \rho_{nn}  \equiv [\rho_{nn}^{\rm ss} - (\rho_{\rm G})_{nn}]/ (\rho_{\rm G})_{nn}$ as a function of $\epsilon^2$ for various $n$ and check whether $\Delta \rho_{nn} \sim \epsilon^4$ as expected from the analytic continuation.  The result is presented in Fig.~\ref{fig1}(b). As seen in the figure, $\Delta \rho_{nn} $ shows $\epsilon^4$ behavior for small $\epsilon$. This clearly demonstrates that the analytic calculation provides the correct answer.

To check the dependence of the ULE steady state on temperature, we perform the same simulation with higher temperature $\beta = 0.05/\eta$. The result is also presented in in Fig.~\ref{fig1}(a). Different from the case of lower temperature $\beta = 0.5/\eta$, the ULE steady state for $\beta = 0.05/\eta$ coincides with the system Gibbs state. From the similar simulations with other parameter sets (not shown here), we find that $\rho_{nn}^{\rm ss}$ tends to approach $(\rho_{\rm G})_{nn}$ for higher temperature. This indicates that the ULE steady state can be practically equivalent to the MFG state in the higher temperature regime for a certain condition.

\subsubsection{Example 2: spin-boson model}
\label{subsec:spin-boson}

In order to see explicitly the difference between the steady states of the ULE and the MFG state, we consider a simpler model. We consider the 
spin-boson model where the two-level system $H_{\rm S}=(\omega_0/2)\sigma_z$ interacting 
with the bath described by bosonic modes with frequency $\omega_k$: 
$H_{\rm B}=\sum_k \omega_k a^\dag_k a_k$ where $a^\dag_k$ and $a_k$ are bosonic creation and annihilation operators, respectively.
We take the interaction Hamiltonian to be 
\begin{align}
H_{\rm I}=(c_x \sigma_x +c_y \sigma_y +c_z \sigma_z)\sum_k\frac{\lambda_k}{2}(a^\dag_k+a_k),
\end{align}
with the coupling constants $c_i$, $i=x,y,z$ and $\lambda_k$.
Using the spectral function $J(\omega)\equiv\sum_k \lambda^2_k \delta(\omega-\omega_k)$, we can show that 
the Fourier transform $\gamma(\omega)$ of the bath correlation function is equal to
$(\pi/2)J(\omega) (\bar{n}(\omega)+1)$ for $\omega>0$
and $(\pi/2)J(-\omega)\bar{n}(-\omega)$ for $\omega<0$, where
$\bar{n}(\omega)=1/(e^{\beta\omega}-1)$.  We take the Ohmic spectral function
\begin{align}
J(\omega)=J_0 \frac{\omega_D \omega}{\omega^2+\omega^2_D},
\end{align}
where the Debye frequency $\omega_D$ serves as a cutoff. Since $J(-\omega)=-J(\omega)$,
we can write $\gamma(\omega)=(\pi/2) J(\omega) (\bar{n}(\omega)+1)$ for all $\omega$.
From Eq.~(\ref{Sab}), we obtain $S(\omega)$. For the ULE, we have $g(\omega)=\sqrt{\gamma(\omega)}$
from Eq.~(\ref{def:g}) and $f(\omega,\omega^\prime)$ is calculated from Eq.~(\ref{eq:f}). We calculate the steady states
of the Redfield equation and the ULE from Eqs. (\ref{rho2nm}), (\ref{rho2nn}), (\ref{rho2nm_ule})
and (\ref{rho2nn_ule}) (see the detailed expressions for the spin boson model in Appendix~\ref{app:sb}).
The results are plotted in Figs.~\ref{fig:sb} (a) and (b), which clearly shows the difference between the ULE steady state and the MFG state. However, we can see that $\rho_{+-}^{(2)}$ and $\rho_{++}^{(2)}$ of the two vanish in the high temperature regime $\beta \rightarrow 0$. Thus, the ULE steady state approaches the system Gibbs state in the high temperature regime in this model.

\begin{figure}
	\resizebox{0.90\columnwidth}{!}{\includegraphics{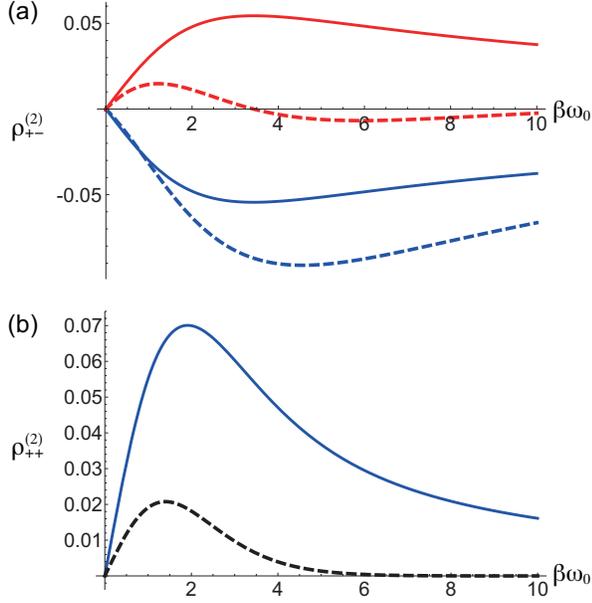}}
	\caption{The second-order correction to the steady-state coherence $\rho^{(2)}_{+-}$ (a) and population $\rho^{(2)}_{++}$ (b) for the spin-boson model described in the text. In the panel (a), the solid blue and red lines are the real and imaginary parts of $\rho^{(2)}_{+-}$ for the Redfield equation or the MFG state, respectively. The dashed blue and red lines are the real and imaginary parts of $\rho^{(2)}_{+-}$  for the ULE, respectively. 
	In the panel (b), the solid line is for the MFG state and the dashed one for the ULE. 
	The parameters used are $c_x=c_y=c_z=J_0=1$, and $\beta\omega_D=10$.  
	}
	\label{fig:sb}
\end{figure}

\subsection{Steady states of the truncated Lindblad equation}

Recently, Becker, Wu, and Eckardt proposed a Lindbladian QME by truncating some part of the Redfield equation~\cite{becker}. 
To be more specific, they decompose the Redfield equation into the Lindblad form and
identify the part that gives the negative contribution, which ruins the complete positivity, and
truncate it. They optimize this process by minimizing this negative contribution \cite{becker}. 
They consider the case that the interaction Hamiltonian is given by $H_{\rm I} = \epsilon A \otimes B $, where $A$ is a dimensionless operator acting on a system and $B$ is an operator with dimension of energy acting on a bath, and both are hermitian. Here we set for $A$ to satisfy a \emph{normalization condition} $\sum_{nm} |A_{nm}|^2 = 1$. Their suggested QME is
\begin{equation}
	\frac{d \rho}{dt} =  - i [ H_{\rm S} + H_{\rm LS}, \rho  ] + \gamma \mathcal{D}_{\rm trc},	\label{eq:truncQME}
\end{equation}
where $H_{\rm LS}$ and $\mathcal{D}_{\rm trc}$ are given by 
\begin{align}
	H_{\rm LS} &= \frac{\epsilon^2}{2 i} (A  \mathbb{A}_t -  \mathbb{A}_t^\dagger A ), \nonumber \\
	\mathcal{D}_{\rm trc} &= L \rho L^\dagger - \frac{1}{2} \{ L^\dagger L, \rho \}.
\end{align}
Here, $L$ and $\mathbb{A}_t$ are
\begin{align}
	L &= \frac{1}{\sqrt{2\cos \varphi_t}} \left( \lambda_t^+ A + \frac{1}{\lambda_t^+} \mathbb{A}_t \right), \nonumber \\ 
	\mathbb{A}_t &= \int_0^t d\tau C(\tau) e^{-iH_{\rm S} \tau} A e^{i H_{\rm S} \tau}
\end{align}
with $C(\tau) = {\rm Tr}_B (\tilde{B}(\tau) B \rho_{\rm B})$ and $\lambda_t^{\pm} = \lambda_t e^{\mp i\varphi_t/2}$. In this formulation, there exists freedom to choose $\lambda_t$ and $\varphi_t$ to minimize the truncated part. Their optimized values are given as $\lambda_t^4 = (\overline{g_t^2} + \overline{h_t^2})^{1/2}$ and $\sin \varphi_t =  \overline{h}_t  / (\overline{g_t^2} + \overline{h_t^2})^{1/2}
$, where the overline denotes an average defined by $\overline{x} = \sum_{nm} x(\Delta_{nm}) |A_{nm}|^2 $ and $g_t = g_t (\Delta_{nm})$ and $h_t = h_t (\Delta_{nm})$ are real and imaginary parts of the function $G_{nm}^t$ defined in Eq.~\eqref{eq:GandW}, respectively.

By using Eqs~\eqref{eqa:Dnm} and \eqref{eqa:H_LS_nm},   in terms of matrix element of $\rho$ based on the eigenstates of $H_{\rm S}$, Eq.~\eqref{eq:truncQME} can be written as
\begin{align}
	\frac{d \rho_{nm}}{dt} &= \langle n| -i [H_{\rm S}, \rho]  | m \rangle +  \langle n| -i [H_{\rm LS}, \rho] + \epsilon^2 \mathcal{D}_{\rm trc}  | m \rangle \nonumber \\
	& \equiv - i \Delta_{nm} \rho_{nm} + \epsilon^2 \sum_{k,l}  \mathcal{R}_{nm,kl}^t \rho_{kl}, \label{eq:truncQME_nm}
\end{align}
where $\Delta_{nm} = E_n - E_m$ and $\mathcal{R}_{nm,kl}^t$ is given by
\begin{align}
	\mathcal{R}_{nm,kl}^t &= -\frac{\delta_{lm}}{2} \sum_j \left[ G_{jk}^t - (G_{jn}^t)^* + (W_{jn}^t)^* W_{jk}^t  \right] A_{nj} A_{jk} \nonumber \\
	& -\frac{\delta_{kn}}{2} \sum_j \left[  (G_{jl}^t)^* -G_{jm}^t+ (W_{jl}^t)^* W_{jm}^t  \right] A_{lj} A_{jm} \nonumber \\
	&+  W_{nk}^t (W_{ml}^t)^* A_{nk} A_{lm}. \label{eq:R_nmkl}
\end{align}
Here $G_{nm}^t$ and $W_{nm}^t$ are defined as
\begin{align}
	G_{nm}^t  &\equiv \int_0^t d\tau e^{-i \Delta_{nm } \tau } C(\tau), \nonumber \\
	W_{nm}^t &\equiv \frac{1}{\sqrt{2 \cos \varphi_t}} \left( \lambda_t^+ + \frac{1}{\lambda_t^+} G_{nm}^t \right),	\label{eq:GandW}
\end{align}
which are also given in Eq.~\eqref{eqa:SS_nm} and \eqref{eqa:A_nm}, respectively. 

Perturbation solution of Eq.~\eqref{eq:truncQME_nm} can be obtained by expanding the density matrix of the system in the steady state ($t=\infty$) up to the order of $\epsilon^2$ as $\rho_{nm}^{\rm ss} = \rho_{nm}^{(0)} + \epsilon^2 \rho_{nm}^{(2)}$. By following the same process as Eqs.~\eqref{st_eq0}, \eqref{st_eq2}, and \eqref{eq_dia}, we have
\begin{align}
	&\rho_{nm}^{(0)} = 0~~~ {\rm   when } ~n \neq m ,  \\
	&\sum_k \mathcal{R}_{nn,kk}^\infty \rho_{kk}^{(0)} = 0 ~~~ {\rm   when } ~n = m , \label{eq:rho_nn_0} \\
	&i \Delta_{nm} \rho_{nm}^{(2)} =  \sum_k \mathcal{R}_{nm,kk}^\infty \rho_{kk}^{(0)}  ~~~ {\rm  when } ~n \neq m. \label{eq:rho_nm_2}
\end{align}
By using Eq.~\eqref{eq:R_nmkl}, Eq.~\eqref{eq:rho_nn_0} can be arranged as
\begin{align}
	0&=-\frac{1}{2} \sum_j \left[ G_{jn}^\infty - (G_{jn}^\infty)^* + (W_{jn}^\infty)^* W_{jn}^\infty  \right] A_{nj} A_{jn} \rho_{nn}^{(0)} \nonumber \\
	&~~~ -\frac{1}{2} \sum_j \left[  (G_{jn}^\infty)^* - G_{jn}^\infty+ (W_{jn}^\infty)^* W_{jn}^\infty  \right] A_{nj} A_{jn} \rho_{nn}^{(0)} \nonumber \\
	&~~~ +  \sum_k W_{nk}^t (W_{nk}^t)^* A_{nk} A_{kn} \rho_{kk}^{(0)} \nonumber \\
	&= \sum_k \left( | W_{nk}^\infty |^2  \rho_{kk}^{(0)} - |W_{kn}^\infty|^2   \rho_{nn}^{(0)}  \right) A_{nk} A_{kn}. \label{eq:first_orderEq} 
\end{align}
From Eqs.~\eqref{eq:GandW} and  \eqref{eq:first_orderEq}, we have
\begin{equation}
	\frac{\rho_{kk}^{(0)}}{\rho_{nn}^{(0)}} = \frac{ |W_{kn}^\infty|^2  }{ |W_{nk}^\infty|^2 } \neq e^{-\beta \Delta_{kn}}.	
\end{equation}
Therefore, the truncated Lindblad equation does not give the system Gibbs state $\rho_{\rm G}$ even in the leading order.

\section{Conclusion}

\begin{table}
\caption{\label{table1} Perturbative steady states of the quantum master equations studied in this paper.
We have also indicated the places where they are calculated. For small coupling constant $\epsilon$
between the system and the bath, the steady state is given by $\rho^{\rm st}=\rho^{(0)}+\epsilon^2\rho^{(2)}+O(\epsilon^4)$.
For each case, the zeroth and the second order contributions are compared with those for the MFG state which is defined in Eq.~(\ref{eq:rG})
and is given by $\rho_{\rm mG}=\rho_{\rm G}+\epsilon^2\rho^{(2)}_{\rm mG}+O(\epsilon^4)$. 
The property of complete positiveness (CP) of each master equation is also shown.
}
\begin{ruledtabular}
\begin{tabular}{llll}
& CP & $\rho^{(0)}$ &  $\rho^{(2)}$\\ \hline 
Redfield (Sec.\ II.A) & No & $\rho_{\rm G}$ & $\rho^{(2)}_{\rm mG}$ \\
Secular Approx.~(Sec.\ III.A) & Yes & $\rho_{\rm G}$ & 0 \\
ULE (Sec.\ III.B)& Yes & $\rho_{\rm G}$ & $\neq\rho^{(2)}_{\rm mG}$\\
TLE (Sec.\ III.C)& Yes & $\neq\rho_{\rm G}$ & $\neq\rho^{(2)}_{\rm mG}$
\end{tabular}
\end{ruledtabular}
\end{table}

In this study, we calculate equilibrium steady states of the recently developed QMEs up to the second order in the coupling strength between a bath and a system of interest. The steady states we have studied in this paper are summarized in Table~\ref{table1}. From this calculation, we explicitly show that the steady states of Lindbladian QMEs differ from the MFG state, which is the equilibrium state when the total system thermalizes. The steady state of the ULE in the leading order is shown to be the system Gibbs state $\rho_{\rm G}$, similar to the Redfield equation. For the ULE, we explicitly evaluate the next-order corrections for the population and coherence, which exhibit discrepancies from those of the MFG state. For the TLE, we find that the steady state differs from the MFG, even in the leading order. Thus, our result clearly demonstrates that any modification of the Redfield equation to achieve complete positivity of a QME can cause undesirable changes to the steady state. However, we also show that this discrepancy can be reduced in the high-temperature regime, where the steady states of the Lindbladian QMEs and MFG can reduce to the same Gibbs state of the system Hamiltonian under certain conditions.  Therefore, the steady-state problem of the Lindblad QMEs can be mitigated by increasing the environmental temperature. Our results lead to another next important question, that is, how can we establish a QME with both complete positivity and a desirable steady state simultaneously for low or moderate temperatures? This requires further study. We anticipate that the recent study on the completely positive QME in zero temperature~\cite{david} can be a good starting point to achieve this goal.

\acknowledgements
We thank Hyunggyu Park for valuable discussions and carefully reading the manuscript. Authors acknowledge the Korea Institute for Advanced Study for providing computing resources 
(KIAS Center for Advanced Computation Linux Cluster System).
This research was supported by NRF Grants No.~2020R1F1A1062833
(J.Y.), and individual KIAS Grants No.~PG064901 (J.S.L.) at the Korea Institute for Advanced Study.

\onecolumngrid
\appendix
\section{Canonical Perturbation Theory}
\label{sec:app_CPT}

Consider the total Hamiltonian in Eq.~\eqref{total} with the interaction Hamiltonian in Eq.~\eqref{interaction}. Using the following identity~\cite{Kubo}
\begin{equation}
	e^{\beta (A+B)} = e^{\beta A}\left[ \mathbb{I} +\int_0^\beta d \lambda e^{-\lambda A} B e^{\lambda (A+B)} \right],
\end{equation}
we have
\begin{align}
	e^{-\beta H} &= e^{-\beta(H_{\rm S}+H_{\rm B})} \left[  \mathbb{I} - \int_0^\beta d \lambda_1 e^{\lambda_1 (H_{\rm S}+H_{\rm B})} H_{\rm I} e^{-\lambda_1 (H_{\rm S}+H_{\rm B})} \right. \nonumber \\
	& ~~~ \left. + \int_0^\beta d \lambda_1 e^{\lambda_1 (H_{\rm S}+H_{\rm B})} H_{\rm I} e^{-\lambda_1 (H_{\rm S}+H_{\rm B})} \int_0^{\lambda_1}  d\lambda_2 e^{\lambda_2 (H_{\rm S}+H_{\rm B})} H_{\rm I} e^{-\lambda_2 (H_{\rm S}+H_{\rm B})}  \right] +O(\epsilon^3) \nonumber \\
	& = e^{-\beta(H_{\rm S}+H_{\rm B})} - \epsilon e^{-\beta H_{\rm S}} \int_0^\beta d \lambda_1 \sum_{\alpha}  \tilde{A}_\alpha (-i \lambda_1)  e^{-\beta H_{\rm B}} \tilde{B}_\alpha (-i \lambda_1) \nonumber \\
	& ~~~+ \epsilon^2 e^{-\beta H_{\rm S}} \int_0^\beta d \lambda_1 \int_0^{\lambda_1} d \lambda_2 \sum_{\alpha,\alpha^\prime} \tilde{A}_\alpha (-i \lambda_1) \tilde{A}_{\alpha^\prime} (-i \lambda_2) e^{-\beta H_{\rm B}} \tilde{B}_\alpha (-i \lambda_1) \tilde{B}_{\alpha'} (-i \lambda_2) +O(\epsilon^3), \label{eqA:eHtot}
\end{align}
where $\tilde{O} (t) \equiv e^{itH_{\rm X}} O e^{-itH_{\rm X}}$ with $\mathrm{X}=\mathrm{S}$ or B. 
Now we take the trace of $e^{-\beta H}$ over the bath degree of freedom. By setting the average of $B_\alpha$ zero, 
i.e. $\langle B_\alpha \rangle_{\rm B} = 0 $, the trace of the second term in Eq~\eqref{eqA:eHtot} vanishes as 
${\rm tr}_{\rm B} \{e^{-\beta H_{\rm B}} \tilde{B}_\alpha (-i \lambda_1) \} =Z_{\rm B} \langle B_\alpha \rangle_{\rm B} =0 $, 
where $Z_B = {\rm tr} (e^{-\beta H_{\rm B}})$. By using the fact ${\rm Tr} \{ e^{-\beta H_{\rm B}} \tilde{B}_\alpha (-i \lambda_1) 
\tilde{B}_{\alpha'} (-i \lambda_2) \} = Z_{\rm B} \langle \tilde{B}_\alpha (-i \lambda_{12}) B_{\alpha'} \rangle_{\rm B} = 
Z_{\rm B} C_{\alpha\alpha'} (-i \lambda_{12})$ with $\lambda_{12}=\lambda_1-\lambda_2$,
the trace of $e^{-\beta H}$ over the bath  becomes
\begin{align}
	{\rm Tr}_{\rm B} \left(e^{-\beta H} \right) = e^{-\beta H_{\rm S}} Z_{\rm B} \left( \mathbb{I}_ {\rm S} + \epsilon^2 \int_0^\beta d \lambda_1 \int_0^{\lambda_1} d \lambda_2 \sum_{\alpha,\alpha'} \tilde{A}_\alpha (-i \lambda_1) \tilde{A}_{\alpha'} (-i \lambda_2) C_{\alpha\alpha'}(-i \lambda_{12}) \right)  +O(\epsilon^3). \label{eqA:traceB_Htot}
\end{align}
Therefore, the trace of Eq.~\eqref{eqA:traceB_Htot} over the  subsystem  is 
\begin{align}
	{\rm tr}_{\rm SB} \left( e^{-\beta H} \right) = Z_{\rm B} Z_{\rm S} + Z_{\rm B}{\rm tr}_{\rm S} (\mathcal{D}) +O(\epsilon^3), \label{eqA:traceTot_Htot}
\end{align} 
where the definition of $\mathcal{D}$ is given in Eq.~\eqref{eq:D}. By using Eqs.~\eqref{eqA:traceB_Htot} and \eqref{eqA:traceTot_Htot},  
the mean force Gibbs state is
\begin{align}
	\rho_{\rm mG} = \frac{{\rm tr}_{\rm B} \left( e^{-\beta H} \right) }{{\rm tr}_{\rm SB} \left( e^{-\beta H} \right)} = \frac{e^{-\beta H_{\rm S}}}{Z_{\rm S}} + \frac{\mathcal{D}}{Z_{\rm S}} - \frac{{\rm tr}_{\rm S} (\mathcal{D})}{Z_{\rm S}^2} e^{-\beta H_{\rm S}} +O(\epsilon^3),
\end{align}
which is the Eq.~\eqref{eq:rho_rG_expansion}. 

We can evaluate the matrix elements $D_{nm}=\langle n\vert \mathcal{D} \vert m\rangle $ as follows.  We change the integration variables as
$x=\lambda_{12}$ and $y=(\lambda_1+\lambda_2)/2$. We can integrate over $y$ to obtain for $n\neq m$
\begin{align}
D_{nm}= \frac{1}{\Delta_{nm}}\sum_{\alpha,\alpha'} \sum_k (A_\alpha)_{nk}  ( A_{\alpha'})_{km} 
\int_0^\beta dx\; C_{\alpha\alpha'}(-i x) \left\{e^{-\beta E_m }e^{x\Delta_{mk}}
-e^{-\beta E_n}e^{x\Delta_{nk}} \right\} .
\label{Dnm}
\end{align} 
The diagonal element is given by
\begin{align}
D_{nn}= e^{-\beta E_n} \sum_{\alpha,\alpha'} \sum_k (A_\alpha)_{nk}  ( A_{\alpha'})_{km} 
\int_0^\beta dx\; (\beta -x ) C_{\alpha\alpha'}(-i x)  e^{x\Delta_{nk}}  .
\label{Dnn}
\end{align}

\begin{figure}
\resizebox{0.3\columnwidth}{!}{\includegraphics{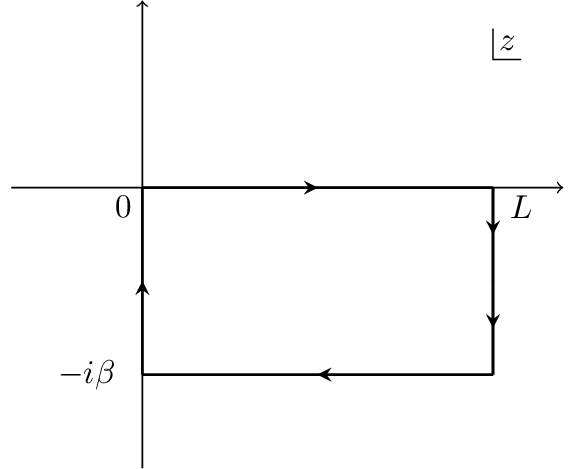}}
\caption{ The contour used in the derivation of Eq.~(\ref{result:mfg}) }
\label{fig:mfg}
\end{figure}

The above integrals can be related to  $S_{\alpha\beta}$ defined in Eq.~(\ref{Sab}) in the following way. We consider 
an integral $\oint dz e^{i\Delta z}C_{\alpha\alpha'}(z)$ along the closed contour shown in Fig.~\ref{fig:mfg}, which vanishes if we assume the analyticity of 
$C_{\alpha\alpha'}(z)$. Since the integral along the right vertical contour vanishes 
in the limit $L\to\infty$, we have
\begin{align}
\int_0^\infty ds\; e^{i\Delta s} C_{\alpha\alpha'} -e^{\beta\Delta}\int_0^\infty ds\; e^{i\Delta s}C^*_{\alpha\alpha'}(s)
+i\int_0^\beta dx\; e^{\Delta x} C_{\alpha\alpha'}(-ix)=0,
\end{align}
where we have used $C_{\alpha\alpha'}(s-i\beta)=C_{\alpha'\alpha}(-s)=C^*_{\alpha\alpha'}(s)$.
Using the definition $\int_0^\infty ds\; e^{i\Delta s} C_{\alpha\alpha'}(s)=(1/2)\gamma_{\alpha\alpha'}(\Delta) +iS_{\alpha\alpha'}(\Delta)$
and the KMS relation $\gamma_{\alpha\alpha'}(\Delta)=e^{\beta\Delta}\gamma_{\alpha'\alpha}(-\Delta)$, we obtain
\begin{align}
\int_0^\beta dx\; e^{\Delta x} C_{\alpha\alpha'}(-ix)=-S_{\alpha\alpha'}(\Delta)-e^{\beta\Delta}S_{\alpha'\alpha}(-\Delta).
\label{result:mfg}
\end{align}
Inserting this into Eq.~(\ref{Dnm}), we can show that $D_{nm}/Z_{\rm S}=\rho^{(2)}_{nm}$ in Eq.~(\ref{rho2nm}) of the Redfield equation.
Taking a derivative of Eq.~(\ref{result:mfg}) with respect to $\Delta$, we obtain
\begin{align}
\int_0^\beta dx\; e^{\Delta x} (\beta-x) C_{\alpha\alpha'}(-ix)=S^\prime_{\alpha\alpha'}(\Delta)-\beta S_{\alpha\alpha'}(\Delta)
-e^{\beta\Delta}S^\prime_{\alpha'\alpha}(-\Delta).
\end{align}
Inserting this into Eq.~(\ref{Dnn}), we find that $D_{nn}/Z_{\rm S}=\bar{\rho}^{(2)}_{nn}$ in Eq.~(\ref{rho2nn}). We have established 
the fact that the second order expansion of the MFG state coincides with the correction to the population and coherence of the 
steady state of the Redfield equation. The high and low temperature limits of the above expressions
have been studied in detail in Ref.~\cite{cresser}.

\section{Demonstration that $\rho_{\rm G}$ is not a steady state of the ULE}
\label{sec:app1}
Using the relations $A_\alpha^\dagger (\omega) = A_\alpha(-\omega)$,
$\rho_\textrm{G} A_\alpha(\omega) = e^{\beta \omega} A_\alpha(\omega) \rho_\textrm{G}$ and
$\rho_\textrm{G} A_\alpha^\dagger(\omega) = e^{-\beta \omega} A_\alpha^\dagger (\omega) \rho_\textrm{G}$, we have
\begin{align}
\mathcal{L}_{\rm ULE}^{(a)}[\rho_{\rm G}]  &=-i \sum_{\alpha,\beta}
 \sum_{\omega,\omega'} f_{\alpha\beta}(-\omega,\omega') \left( A^\dag_\alpha(\omega) A_\beta(\omega) \rho_\textrm{G}
 - \rho_\textrm{G} A^\dag_\alpha(\omega) A_\beta(\omega)\right) \\
&=-i\sum_{\alpha,\beta}
 \sum_{\omega,\omega'} f_{\alpha\beta}(-\omega,\omega')  \left(1-e^{-\beta(\omega-\omega')}\right)A^\dag_\alpha(\omega)
  A_\beta(\omega')  \rho_\textrm{G}.
\label{eq:result1}
\end{align}
For the dissipator part, we have
\begin{align}
\sum_\alpha L_\alpha \rho_\textrm{th} L^\dagger_\alpha =&\sum_{\alpha,\beta,\gamma}
\sum_{\omega,\omega'} g_{\alpha\beta}(\omega)g^*_{\alpha\gamma}(\omega')A_\beta(\omega) \rho_{\rm G}A^\dag_\gamma(\omega') \nonumber \\
=&\sum_{\alpha,\beta,\gamma}
\sum_{\omega,\omega'} g_{\alpha\beta}(-\omega)g_{\gamma\alpha}(-\omega')A^\dag_\beta(\omega)A_\gamma(\omega') \rho_{\rm G} e^{\beta\omega'},
\end{align}
and
\begin{align}  
  \left\{ \sum_\alpha L_\alpha^\dagger L_\alpha, \rho_\textrm{th} \right\}
=& \sum_{\alpha,\beta,\gamma}
\sum_{\omega,\omega'} g^*_{\alpha\beta}(\omega)g_{\alpha\gamma}(\omega')\left(A^\dag_\beta(\omega) A_\gamma(\omega') \rho_{\rm G}
+\rho_{\rm G}A^\dag_\beta(\omega) A_\gamma(\omega')\right)\nonumber \\
=&\sum_{\alpha,\beta,\gamma}
\sum_{\omega,\omega'} g_{\beta\alpha}(\omega)g_{\alpha\gamma}(\omega')\left(1+e^{-\beta(\omega-\omega')}\right)
A^\dag_\beta(\omega) A_\gamma(\omega') \rho_{\rm G} .
 \label{eq:result}
\end{align}
Using the KMS relation $g_{\alpha\beta}(-\omega)=e^{-\beta\omega/2}g_{\beta\alpha}(\omega)$, we have
\begin{align}
\mathcal{L}^{(b)}_{\rm ULE}[\rho_{\rm G}]=-\frac 1 2 \sum_{\alpha,\beta,\gamma}
\sum_{\omega,\omega'} g_{\beta\alpha}(\omega)g_{\alpha\gamma}(\omega')\left(1-e^{-\beta(\omega-\omega')}\right)^2
A^\dag_\beta(\omega) A_\gamma(\omega') \rho_{\rm G} 
\label{eq:result2}
\end{align}
It is clear that Eqs.~(\ref{eq:result1}) and (\ref{eq:result2}) do not vanish in general
due to the presence of the off-diagonal terms in the double frequency sums. 

\section{Perturbative steady state of the ULE}
\label{sec:app2}

We first write the matrix elements of the operators in Eqs.~(\ref{eq:Lambda}) and (\ref{eq:L}), respectively as
\begin{align}
\Lambda_{nm}=\sum_{\alpha,\beta}\sum_k f_{\alpha,\beta}(\Delta_{kn},\Delta_{mk})(A_\alpha)_{nk}(A_\beta)_{km},
\end{align}
and
\begin{align}
(L_\alpha)_{nm}=\sum_\beta g_{\alpha\beta}(\Delta_{mn})(A_\beta)_{nm}.
\end{align}
From Eq.~(\ref{ule_l2}), we have the matrix elements of the superoperators as
\begin{align}
\left(\mathcal{L}^{(a)}_{\rm ULE}[\rho(t)]\right)_{nm}=&-i\sum_{\alpha,\beta}\sum_{k,l}\left[ 
f_{\alpha\beta}(\Delta_{ln},\Delta_{kl}) (A_\alpha)_{nl}(A_\beta)_{lk}\rho_{km} - 
f_{\alpha\beta}(\Delta_{lk},\Delta_{ml}) (A_\alpha)_{kl}(A_\beta)_{lm}\rho_{nk}\right] \nonumber \\
=&-i \sum_{\alpha,\beta}\sum_{j,k,l}\left[
\delta_{lm} f_{\alpha\beta}(\Delta_{jn},\Delta_{kj}) (A_\alpha)_{nj}(A_\beta)_{jk} 
-\delta_{kn} f_{\alpha\beta}(\Delta_{jl},\Delta_{mj}) (A_\alpha)_{lj}(A_\beta)_{jm}\right] \rho_{kl}.
\label{L2_a_ule}
\end{align}
For the dissipator part, we first calculate using the hermiticity of $g_{\alpha\beta}$ 
\begin{align}
\langle n\vert \sum_\alpha L_\alpha \rho(t) L^\dag_\alpha \vert m \rangle = \sum_{\alpha,\beta,\gamma}\sum_{k,l}
g_{\alpha\beta}(\Delta_{kn})g_{\gamma\alpha}(\Delta_{lm})(A_\beta)_{nk}(A_\gamma)_{lm}\rho_{kl},
\end{align}
and
\begin{align}
\langle n\vert \{ \sum_\alpha L^\dag _\alpha L_\alpha , \rho(t) \} \vert m \rangle = &\sum_{\alpha,\beta,\gamma}\Big[ \sum_{j,k}
g_{\beta\alpha}(\Delta_{nj})g_{\alpha\gamma}(\Delta_{kj})(A_\beta)_{nj}(A_\gamma)_{jk}\rho_{km} \nonumber \\
&~~~~~+\sum_{j,l}g_{\beta\alpha}(\Delta_{lj})g_{\alpha\gamma}(\Delta_{mj})(A_\beta)_{lj}(A_\gamma)_{jm}\rho_{nl}\Big].
\end{align}
We therefore have
\begin{align}
\left(\mathcal{L}^{(b)}_{\rm ULE}[\rho(t)]\right)_{nm}= \sum_{\alpha,\beta,\gamma}\sum_{k,l} & \Bigg[ 
g_{\alpha\beta}(\Delta_{kn})g_{\gamma\alpha}(\Delta_{lm})(A_\beta)_{nk}(A_\gamma)_{lm} \nonumber \\
&-\frac 1 2 \delta_{lm} g_{\beta\alpha}(\Delta_{nj})g_{\alpha\gamma}(\Delta_{kj})(A_\beta)_{nj}(A_\gamma)_{jk} \nonumber\\
&-\frac 1 2 \delta_{kn} g_{\beta\alpha}(\Delta_{lj})g_{\alpha\gamma}(\Delta_{mj})(A_\beta)_{lj}(A_\gamma)_{jm}\Bigg]\rho_{kl}.
\label{L2_b_ule}
\end{align}

If we insert $\rho_{kl}=\delta_{kl}\rho^{(0)}_{kk}$ in Eqs.~(\ref{L2_a_ule}) and (\ref{L2_b_ule}), we have, respectively, 
\begin{align} 
\left(\mathcal{L}^{(a)}_{\rm ULE}[\rho^{(0)}]\right)_{nm}=-i \sum_{\alpha,\beta}\sum_{k}
f_{\alpha\beta}(\Delta_{kn},\Delta_{mk}) (A_\alpha)_{nk}(A_\beta)_{km} \left(\rho^{(0)}_{mm}-\rho^{(0)}_{nn}\right),
\end{align}
and
\begin{align}
\left(\mathcal{L}^{(b)}_{\rm ULE}[\rho^{(0)}]\right)_{nm}= \sum_{\alpha,\beta,\gamma}\sum_{k} 
(A_\beta)_{nk}&(A_\gamma)_{km} \Big[ 
g_{\alpha\beta}(\Delta_{kn})g_{\gamma\alpha}(\Delta_{km}) \rho^{(0)}_{kk} \nonumber \\
&-\frac 1 2 g_{\beta\alpha}(\Delta_{nk})g_{\alpha\gamma}(\Delta_{mk})\left(\rho^{(0)}_{mm}+\rho^{(0)}_{nn}\right) \Big].
\end{align}
If we consider only the diagonal elements, we have
\begin{align}
\left(\mathcal{L}^{(a)}_{\rm ULE}[\rho^{(0)}]\right)_{nn}=0,
\end{align}
and
\begin{align}
\left(\mathcal{L}^{(b)}_{\rm ULE}[\rho^{(0)}]\right)_{nn}=\sum_{\alpha,\beta,\gamma}&\sum_{k}\big[ g_{\alpha\beta}(\Delta_{kn})
g_{\gamma\alpha}(\Delta_{kn})\rho^{(0)}_{kk} -g_{\beta\alpha}(\Delta_{nk})g_{\alpha\gamma}(\Delta_{nk})\rho^{(0)}_{nn}\big]
(A_\beta)_{nk}(A_{\gamma})_{kn}.
\end{align}
This gives the condition in Eq.~(\ref{cond1_ule}).

\section{Steady states of the spin boson model}
\label{app:sb}

The Hamiltonian of the spin boson model defined in Sec.~\ref{subsec:spin-boson} is
\begin{align}
H&=H_{\rm S}+H_{\rm B}+H_{\rm I} \nonumber \\
&=\frac{\omega_0}2\sigma_z+\sum^\infty_{k=1} \omega_k a^\dag_k a_k+
(c_x \sigma_x +c_y \sigma_y +c_z \sigma_z)\sum^\infty_{k=1}\frac{\lambda_k}{2}(a^\dag_k+a_k)
\end{align}
Since $\tilde{B}(s)=\sum_k (\lambda_k/2) (e^{i\omega_k s}a^\dag_k +e^{-i\omega_k s}a_k)$, the bath correlation function
$C(t)=\mathrm{Tr}(\tilde{B}(t) B \rho_{\rm B})$ is given by
\begin{align}
C(t)=&\frac 1 4 \int_0^\infty d\omega J(\omega) \left[e^{i\omega t}\bar{n}(\omega)+e^{-i\omega t}(\bar{n}(\omega)+1)\right] \nonumber \\
=&\frac 1 4 \int_0^\infty d\omega J(\omega) \left[\cos(\omega t)\coth(\frac{\beta\omega}{2})-i\sin(\omega t)\right],
\end{align} 
with the spectral function $J(\omega)=\sum^\infty_{k=1} \lambda^2_k \delta(\omega-\omega_k)$
and the mean occupation number $\bar{n}(\omega)=1/(e^{\beta\omega} -1)$.
In the high temperature limit, the correlation function goes like $\beta^{-1}$, which satisfies the condition, Eq.~\eqref{eq:rho_rG_highT}, for the MFG state as discussed in Sec.~\ref{subsec:mfg}. 
We consider the Ohmic spectral function
\begin{align}
J(\omega)=J_0 \frac{\omega_D \omega}{\omega^2+\omega^2_D},
\end{align}
with he Debye frequency $\omega_D$.
As discussed in the main text, the Fourier transform of $C(t)$ is given by
\begin{align}
\gamma(\omega)=\frac{\pi}2 J(\omega) (\bar{n}(\omega)+1).
\end{align} 

The zeroth order steady state is the system Gibbs state with $\rho^{\rm G}_{++}=e^{-\beta\omega_0/2}/Z_{\rm S}$ and
$\rho^{\rm G}_{--}=e^{\beta\omega_0/2}/Z_{\rm S}$ with $Z_{\rm S}=e^{-\beta\omega_0/2}+
e^{\beta\omega_0/2}$. The components of the system operator are $A_{++}=-A_{--}=c_z$
and $A_{+-} =A^*_{-+}=c_x-ic_y$. 
The perturbative steady state of the Redfield equation, or the second order expansion of MFG state, is given by 
Eqs.~(\ref{rho2nm}) and Eq.~(\ref{rho2nn}). We have 
\begin{align}
(\rho^{(2)}_{\rm Red})_{+-}=&\frac{2c_z(c_x-ic_y)}{\omega_0}\left[ S(0)-\rho^{\rm G}_{++}S(\omega_0)-
\rho^{\rm G}_{--}S(-\omega_0)\right], \\
(\bar{\rho}^{(2)}_{\rm Red})_{++}=& c^2_z [-\beta S(0)]\rho^{\rm G}_{++} +(c^2_x+c^2_y)\left[ (S^\prime(\omega_0)-\beta S(\omega_0))\rho^{\rm G}_{++}
-S^\prime(-\omega_0)\rho^{\rm G}_{--}\right], \\
(\bar{\rho}^{(2)}_{\rm Red})_{--}=&c^2_z [-\beta S(0)]\rho^{\rm G}_{--} +(c^2_x+c^2_y)\left[ (S^\prime(-\omega_0)-\beta S(-\omega_0))\rho^{\rm G}_{--}
-S^\prime(\omega_0)\rho^{\rm G}_{++}\right], 
\end{align}
where $S(\omega)$ and its derivative are obtained from evaluating the principal value integral
\begin{align}
	S(\omega)=-\mathcal{P}\int^\infty_{-\infty}\frac{d\tilde{\omega}}{2\pi}\frac{ \gamma(\tilde{\omega}+\omega)} {\tilde{\omega}} .
\end{align}
From the normalization, we have the correction to the population as
\begin{align}
(\rho^{(2)}_{\rm Red})_{++}=(\bar{\rho}^{(2)}_{\rm Red})_{++}-\rho^{\rm G}_{++} \left[(\bar{\rho}^{(2)}_{\rm Red})_{++}+(\bar{\rho}^{(2)}_{\rm Red})_{--}\right]
\end{align}
and $(\rho^{(2)}_{\rm Red})_{--}=-(\rho^{(2)}_{\rm Red})_{++}$.

For the ULE, we first note that $g(\omega)=\sqrt{\gamma(\omega)}$. From Eq.~(\ref{rho2nm_ule}), we obtain
\begin{align}
(\rho^{(2)}_{\rm ULE})_{+-}=\frac{c_z(c_x-ic_y)}{\omega_0}&\Big[ \{f(0,-\omega_0)-f(-\omega_0,0)\}
\{ \rho^{\rm G}_{++} -\rho^{\rm G}_{--} \} \nonumber \\
+&\frac{i}{2}g(0) \left[ \{g(-\omega_0)-3g(\omega_0)\}\rho^{\rm G}_{++}
-\{ g(\omega_0)-3 g(-\omega_0)\}\rho^{\rm G}_{--}\right]\Big],
\end{align}
where
\begin{align}
f(0,-\omega_0) =  f(\omega_0,0) =-\mathcal{P}\int^\infty_{-\infty}
\frac{d\tilde{\omega}}{2\pi}\frac{ g(\tilde{\omega})g(\tilde{\omega}-\omega_0)}{\tilde{\omega}}
\end{align}
from Eq.~(\ref{eq:f}). 
The diagonal part is given from Eq.~(\ref{rho2nn_ule}) as
\begin{align}
(\bar{\rho}^{(2)}_{\rm ULE})_{++}=& -\beta \rho^{\rm G}_{++}\left[  c^2_z S(0)  +(c^2_x+c^2_y) S(\omega_0)
\right] , \\
(\bar{\rho}^{(2)}_{\rm ULE})_{--}=&-\beta \rho^{\rm G}_{--}\left[  c^2_z S(0)  +(c^2_x+c^2_y) S(-\omega_0)
\right] ,
\end{align}
where we have used $f(0,0)=S(0)$ and $f(-\omega_0,\omega_0)=S(\omega_0)$. For the ULE, the correction to population is then
\begin{align}
(\rho^{(2)}_{\rm ULE})_{++}=(\bar{\rho}^{(2)}_{\rm ULE})_{++}-\rho^{\rm G}_{++} \left[(\bar{\rho}^{(2)}_{\rm ULE})_{++}+(\bar{\rho}^{(2)}_{\rm ULE})_{--}\right]
\end{align}
and $(\rho^{(2)}_{\rm ULE})_{--}=-(\rho^{(2)}_{\rm ULE})_{++}$.

\section{Matrix elements of Eq.~\eqref{eq:truncQME_nm} }
\label{sec:app3}
Here, we evaluate the matrix elements related to Eq.~\eqref{eq:truncQME_nm} in terms of the basis of $H_{\rm S}$. First, $\langle n| \mathbb{A}_t | m \rangle$ is  
\begin{equation}
	\langle n| \mathbb{A}_t | m \rangle = \int_0^t d \tau C(\tau) e^{-i \Delta_{nm} \tau} A_{nm} = G_{nm}^t	A_{nm},~~{\rm where~} G_{nm}^t  \equiv \int_0^t d\tau e^{-i \Delta_{nm } \tau } C(\tau). \label{eqa:SS_nm}
\end{equation}
Thus, $\langle n| \mathbb{A}_t^\dagger | m \rangle = \langle m| \mathbb{A}_t | n \rangle^* = (G_{mn}^t)^* A_{nm} $, where the last equality comes from the hermitian property $A_{mn}^* = A_{nm}$. Next, $\langle n| L | m \rangle$ becomes
\begin{align}
	L_{nm} = \langle n| L | m \rangle = \frac{1}{\sqrt{2 \cos \varphi_t}} \left( \lambda_t^+ A_{nm} + \frac{1}{\lambda_t^+} G_{nm}^t A_{nm}  \right) \equiv W_{nm}^t A_{nm},~~{\rm where~} W_{nm}^t \equiv \frac{1}{\sqrt{2 \cos \varphi_t}} \left( \lambda_t^+ + \frac{1}{\lambda_t^+} G_{nm}^t \right). \label{eqa:A_nm}
\end{align}
Thus, $(L^\dagger)_{nm} = L_{mn}^* = (W_{mn}^t )^* A_{nm}$.
To evaluate the matrix element of $\mathcal{D}_{\rm trc}$, it is necessary to calculate $\langle n| L \rho L^\dagger | m \rangle$ and $\langle n| \{ L^\dagger L , \rho \} | m \rangle$.
Using Eq.~\eqref{eqa:A_nm}, we have
\begin{align}
	\langle n| L \rho L^\dagger | m \rangle &= \sum_{k,l} L_{nk} \rho_{kl} (L^\dagger)_{lm} = \sum_{k,l} W_{nk}^t (W_{ml}^t)^* A_{nk} A_{lm} \rho_{kl}, \nonumber \\
	\langle n| \{ L^\dagger L , \rho \} | m \rangle &= \sum_{k,l} \left[ (L^\dagger)_{nk} L_{kl} \rho_{lm} + \rho_{nk} (L^\dagger)_{kl} L_{lm}  \right] \nonumber \\ 
	&= \sum_{k,l} \left[ (W_{kn}^t)^* W_{kl}^t A_{nk} A_{kl} \rho_{lm} + (W_{lk}^t)^* W_{lm}^t A_{kl} A_{lm} \rho_{nk}  \right] \nonumber \\
	&= \sum_{j,k}  (W_{jn}^t)^* W_{jk}^t A_{nj} A_{jk} \rho_{km} + \sum_{j,l} (W_{jl}^t)^* W_{jm}^t A_{lj} A_{jm} \rho_{nl}   \nonumber \\
	&= \sum_{k,l} \left[ \delta_{lm} \sum_j (W_{jn}^t)^* W_{jk}^t A_{nj} A_{jk} + \delta_{kn} \sum_j (W_{jl}^t)^* W_{jm}^t A_{lj} A_{jm} \right] \rho_{kl}. \label{eqa:ArA_nm}
\end{align}
From Eq.~\eqref{eqa:ArA_nm}, $\langle n| \mathcal{D}_{\rm trc} | m \rangle$ becomes
\begin{align}
	\langle n| \mathcal{D}_{\rm trc} | m \rangle = \sum_{k,l} \left[ W_{nk}^t (W_{ml}^t)^* A_{nk} A_{lm} 
	-\frac{1}{2}  \delta_{lm} \sum_j (W_{jn}^t)^* W_{jk}^t A_{nj} A_{jk} -\frac{1}{2} \delta_{kn} \sum_j (W_{jl}^t)^* W_{jm}^t A_{lj} A_{jm}  \right]  \rho_{kl}. \label{eqa:Dnm}
\end{align}
The matrix element of $(H_{\rm LS})_{nm}= \langle n| H_{\rm LS} | m \rangle$ is given by
\begin{align}
	(H_{\rm LS})_{nm} = \frac{1}{2 i} \sum_k \left[ A_{nk} \mathbb{A}_{km}^t - (\mathbb{A}_t^\dagger)_{nk} A_{km} \right] = \frac{1}{2 i } \sum_k \left[ G_{km}^t - (G_{kn}^t)^*  \right] A_{nk} A_{km}.
\end{align}
Then, the commutator $ -i\langle n| [H_{\rm LS}, \rho] | \rangle$ is evaluated as
\begin{align}
	-i\langle n| [H_{\rm LS}, \rho] | m\rangle &= -\frac{1}{2} \sum_k \left[ \sum_l \left\{ G_{lk}^t - (G_{ln}^t)^* \right\} A_{nl} A_{lk} \rho_{km} - \sum_l \left\{ G_{lm}^t -(G_{lk}^t)^* \right\} A_{kl} A_{lm} \rho_{nk}   \right] \nonumber \\
	&= -\frac{1}{2} \sum_{k,l} \left[ \delta_{lm} \sum_j \left\{ G_{jk}^t - (G_{jn}^t)^* \right\} A_{nj} A_{jk} \rho_{kl} - \delta_{kn} \sum_j \left\{ G_{jm}^t -(G_{jl}^t)^* \right\} A_{lj} A_{jm} \rho_{kl}   \right]. \label{eqa:H_LS_nm}
\end{align}

\twocolumngrid

\begin{thebibliography}{}
\bibitem{g_convention} Our $g_{\alpha\beta}(\omega)$ is equal to $2\pi$ times the same quantity used in Ref.~\cite{ule}.
\end{thebibliography}
{}

\end{document}